\documentclass[showpacs,aps,twocolumn]{revtex4}

\usepackage{bm}
\usepackage{amsmath}
\usepackage{graphicx}
\usepackage{subfigure}
\usepackage[usenames,dvipsnames]{color}
\definecolor{darkblue}{RGB}{0,0,196}
\usepackage[colorlinks=true,linkcolor=darkblue,citecolor=darkblue,urlcolor=darkblue]{hyperref}

\usepackage{xcolor}

\usepackage{setspace}
\usepackage{footmisc}
\usepackage[makeroom]{cancel}
\usepackage{comment}
\usepackage{lineno}

\def\be{\begin{equation}}
\def\ee{\end{equation}}
\def\ba{\begin{eqnarray}}
\def\ea{\end{eqnarray}}

\usepackage{graphicx}
\usepackage{amsmath,bbm}
\usepackage{amssymb,bm}
\usepackage{yfonts}

\begin{document}
 
\title{ Exploring the effect of hadron cascade-time on particle production in Xe+Xe collisions at $\sqrt{s_{\rm{NN}}}$ = 5.44 TeV using a multi-phase transport model}
\author{Girija Sankar Pradhan$^{1}$}
\author{Rutuparna Rath$^{1}$}
\author{Ronald Scaria$^{1}$}
\author{Raghunath Sahoo$^{1,2,}$\footnote{Corresponding author: $Raghunath.Sahoo@cern.ch$}}
\affiliation{$^{1}$Department of Physics, Indian Institute of Technology Indore, Simrol, Indore 453552, India}
\affiliation{$^{2}$CERN, CH 1211, Geneva 23, Switzerland}

\begin{abstract}

Heavy-ion collisions at ultra-relativistic energies provide extreme conditions of energy density and temperature to produce a deconfined state of quarks and gluons. Xenon (Xe) being a deformed nucleus further gives access to the effect of initial geometry on final state particle production. This study focuses on the effect of nuclear deformation and hadron cascade-time on the particle production and elliptic flow using  A Multi-Phase Transport (AMPT) model in Xe+Xe collisions at $\sqrt{s_{\rm NN}}$ = 5.44 TeV. 
We explore the effect of hadronic cascade-time on identified particle production through the study of $p_{\rm T}$-differential particle ratios. 
The effect of hadronic cascade-time on the generation of elliptic flow is studied by varying the cascade-time between 5 and 25 fm/$c$. This study shows the final state interactions among particles generate additional anisotropic flow with increasing hadron cascade-time especially at very low and high-$p_{\rm T}$. 
\end{abstract}
 
\date{\today}
\maketitle 
\section{Introduction}
\label{intro}

The primary goal of ultra-relativistic heavy-ion collisions is to create a system of deconfined quarks and gluons known as quark-gluon-plasma (QGP) under extreme conditions of high temperature and/or baryon density, and characterise its properties. It is very important to understand the particle production mechanism in ultra-relativistic heavy-ion collisions for which several experimental set-ups are designed at places such as Relativistic Heavy-ion Collider (RHIC) at BNL, USA and Large Hadron Collider (LHC) at CERN, Switzerland. In such collisions, huge initial energy density and pressure are created which drive the expansion of the system through various complex processes
involving different degrees of freedom of the system. During this expansion,  multiple interactions of quarks and gluons may lead to a  thermalized system that further undergoes collective expansion and hadronization to form composite hadrons from the deconfined partons. The hadronized final state particles carry information about the initial-state effects of the system produced during heavy-ion collisions. The initial state geometry of produced system gives rise to spatial anisotropy which later results in the momentum anisotropy of the final state particles in non-central collisions.  Anisotropic flow quantifies the momentum anisotropy of the produced system, which is one of the signatures of QGP. The strength of this anisotropic flow can be estimated from the flow coefficients ($v_{\rm n}$) in the Fourier expansion of the momentum distribution of the final state particles given by,

\begin{equation}
E \frac{d^3N}{dp^3} = \frac{d^2N}{2\pi p_{\rm T} dp_{\rm T} dy} \bigg[1+2\sum_{n=1}^{\infty} v_{n}~ cos[n(\varphi-\psi_{n})]\bigg].
\end{equation}

Where, the second-order Fourier coefficient of the anisotropic flow is called elliptic flow ($v_2$), which is sensitive to the equation of state (EoS) and transport properties \cite {Snellings:2011sz}.  Here $\varphi$ is the azimuthal angle of emission of a final state particle and $\psi_n$ is the angle with respect to the reaction plane-- the plane created by the impact parameter and beam axis. The elliptic flow ($v_2$) has been studied extensively using different colliding systems such as Au+Au, Cu+Cu, U+U, Xe+Xe and Pb+Pb by RHIC and the LHC with a wide range of collision energies varying from 7.7 GeV to 5.44 TeV \cite{Feng:2016emh, v2_phenix3, v2_phenix4, AMPT_Scaling, v2_star2, AMPT_PbPb, Tripathy:2018rln}. The shape of the colliding nuclei may also contribute to the initial state geometry of the produced system, which may subsequently affect the elliptic flow. A comparison of $v_{2}$ between spherical nuclei and a deformed one in the central collisions can establish the origin of elliptic flow due to the initial state effect. Recently, LHC has collided intrinsically deformed nuclei Xenon (Xe) at $\sqrt{s_{\rm NN}}$ = 5.44 TeV, which also bridges the final state charged particle multiplicity gap between the smaller systems (p+p and p+Pb) and larger system (Pb+Pb). For the most central collisions  (0--5$\%$), the $v_{2}$ is found to be $\sim$ 35$\%$ higher in Xe+Xe as compared to Pb+Pb \cite{Acharya:2018ihu}. Further, the violation of quark participant scaling of identified particle $v_{2}$ is also observed for the Xe+Xe system like other colliding systems having spherical nuclei \cite{Kim:2018ink, Tripathy:2018bib}. 

The space-time evolution of the system produced in heavy-ion collisions goes through different stages and the phase between the chemical and kinetic freeze-out boundaries is known as the hadronic phase. At the chemical freeze-out boundary the inelastic processes cease and the particle abundances are fixed which further experience elastic collisions with other particles present in the system till they arrive at the kinetic freeze-out boundary. Inside the hadronic phase, the strength of the scatterings depends on the cross-section of the system constituents and the lifetime of the hadronic phase. The momentum distribution of final state hadrons may also vary due to multiple scatterings in the hadronic phase. This may also contribute to the momentum anisotropy of the final state particles which results in the change in $v_{2}$. This effect can be better studied by varying the lifetime of the hadronic phase, which is possible to carry out with theoretical models like A Multi-Phase Transport model (AMPT).  Earlier, the effect of the hadronic phase on the $v_{2}$ and resonance production has been observed by considering spherical nuclei \cite{Singha:2015fia,Nayak:2017gpi}. It would be interesting to look into the identified and charged particle $v_{2}$ with different hadronic cascade-time ($\tau_{\rm HC}$) as a function of transverse momentum ($p_{\rm T}$) and centrality using the intrinsically deformed nuclei, like Xe.  It should be noted here that the space-time evolution of the produced fireball in heavy-ion collisions is governed by the initial energy density, which is controlled by the available centre-of-mass energy and
the number of participating nuclei. Accordingly, the hadronic phase lifetime and the cascade-time (we use these terms interchangeably
without the violation of the meaning thereof) becomes a function of collision energy and collision geometry of the initial state of the collision.     

The study of particle ratios with different mass, quark content and baryons/mesons can be helpful to understand the particle production mechanism in heavy-ion collisions.
The $p_{\rm T}$-differential particle ratios help to investigate various reasons for particle production at different $p_{\rm T}$ scales \cite{Rath:2018ytr}. At lower $p_{\rm T}$ region, the spectral shapes of identified particles strongly depend on the radial flow that has been observed at the LHC in central heavy-ion collisions which are well explained by hydrodynamically motivated models \cite{Abelev:2013vea,Bozek:2012qs}. The collective expansion of the system pushes the heavier mass particles from low-$p_{\rm T}$  to intermediate-$p_{\rm T}$, which subsequently shows an enhancement of the ratio of the heavier to lighter particles. At intermediate $p_{\rm T}$ region, a clear enhancement of baryon over meson has been observed especially for p/$\pi$ ratios in the most central Pb+Pb collisions as compared to pp-collisions. Baryons to meson ratio are further enhanced by the coalescence mechanism which also requires radial flow where coalescence of three quarks results in the creation of larger $p_{\rm T}$ baryons compared to two quark/antiquark mesons. These $p_{\rm T}$-differential particle ratios can also be affected due to the interactions of identified particles in the hadronic phase through subsequent elastic scatterings. The intensity of such phenomena mostly depends on the hadronic phase lifetime and the elastic cross-section of the identified particles. The high-$p_{\rm T}$ sector is dominated by the jet fragmentation and pQCD processes. 

Here, we present a comprehensive study on the effect of the hadronic phase lifetime (or hadronic cascade-time, as is known otherwise) on the $p_{\rm T}$-differential particle ratios along with $p_{\rm T}$-differential and integrated elliptic flow. In this study, we have used AMPT (string melting version) to generate events at different hadronic phase times varying from 5 to 25 fm/$c$, which will be discussed in detail in Section \ref{formalism}. Further, we have discussed the results in the Section \ref{ResDis}, which is followed by the Section \ref{summary} with a summary of this study.

 \begin{figure*}[hbt!]
\includegraphics[width=6cm, height=6cm]{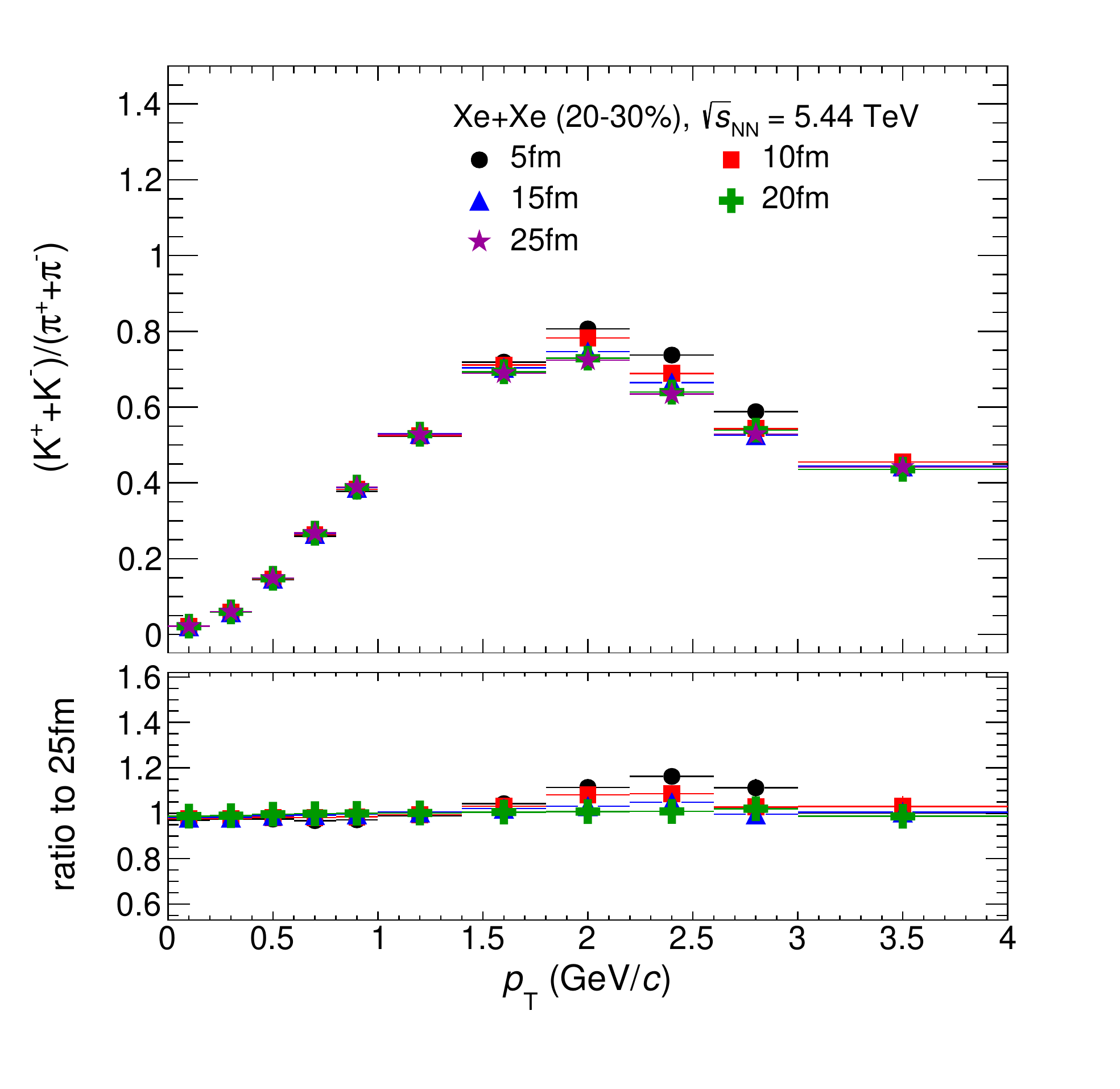}
\includegraphics[width=6cm, height=6cm]{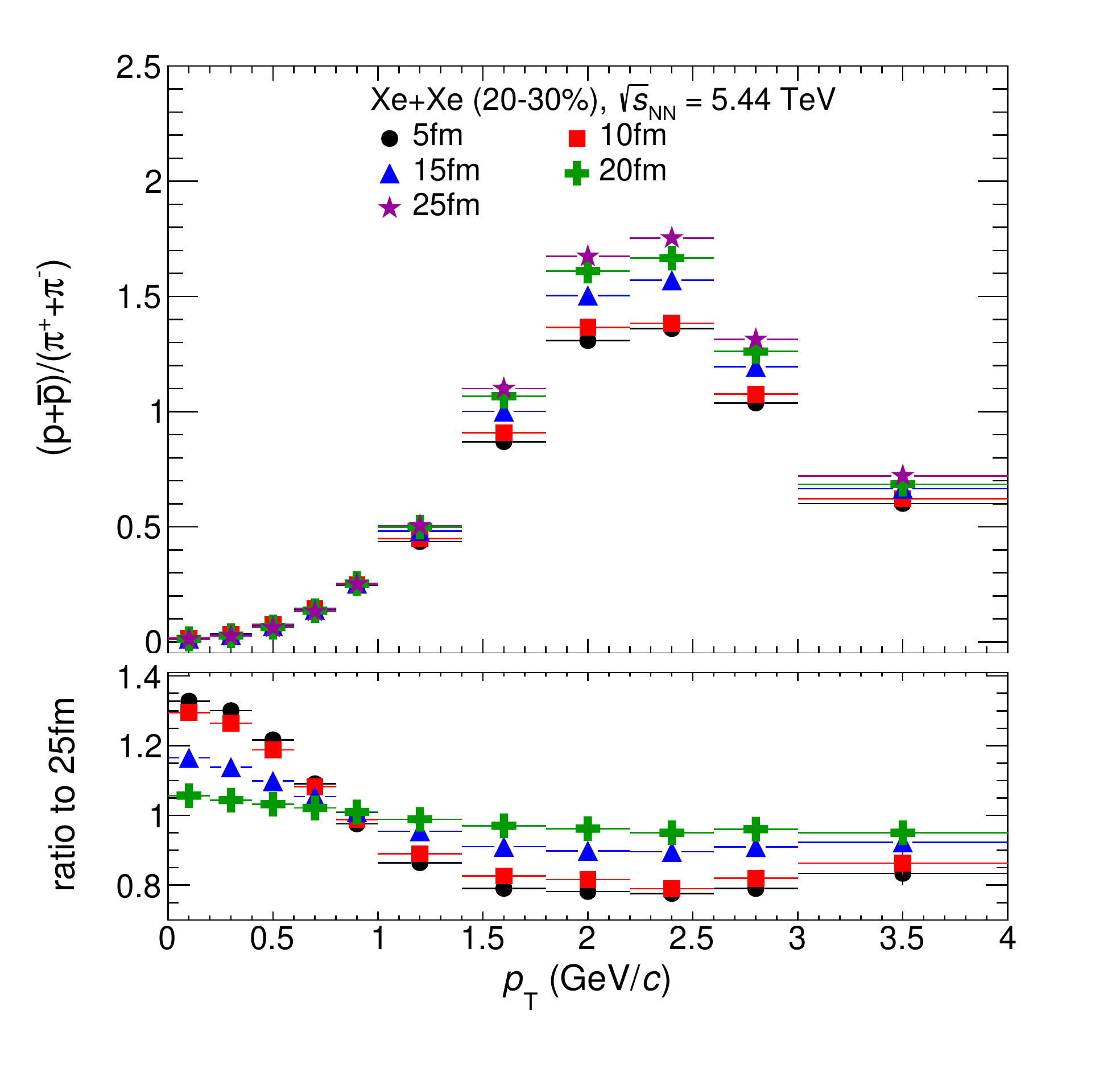}
\includegraphics[width=6cm, height=6cm]{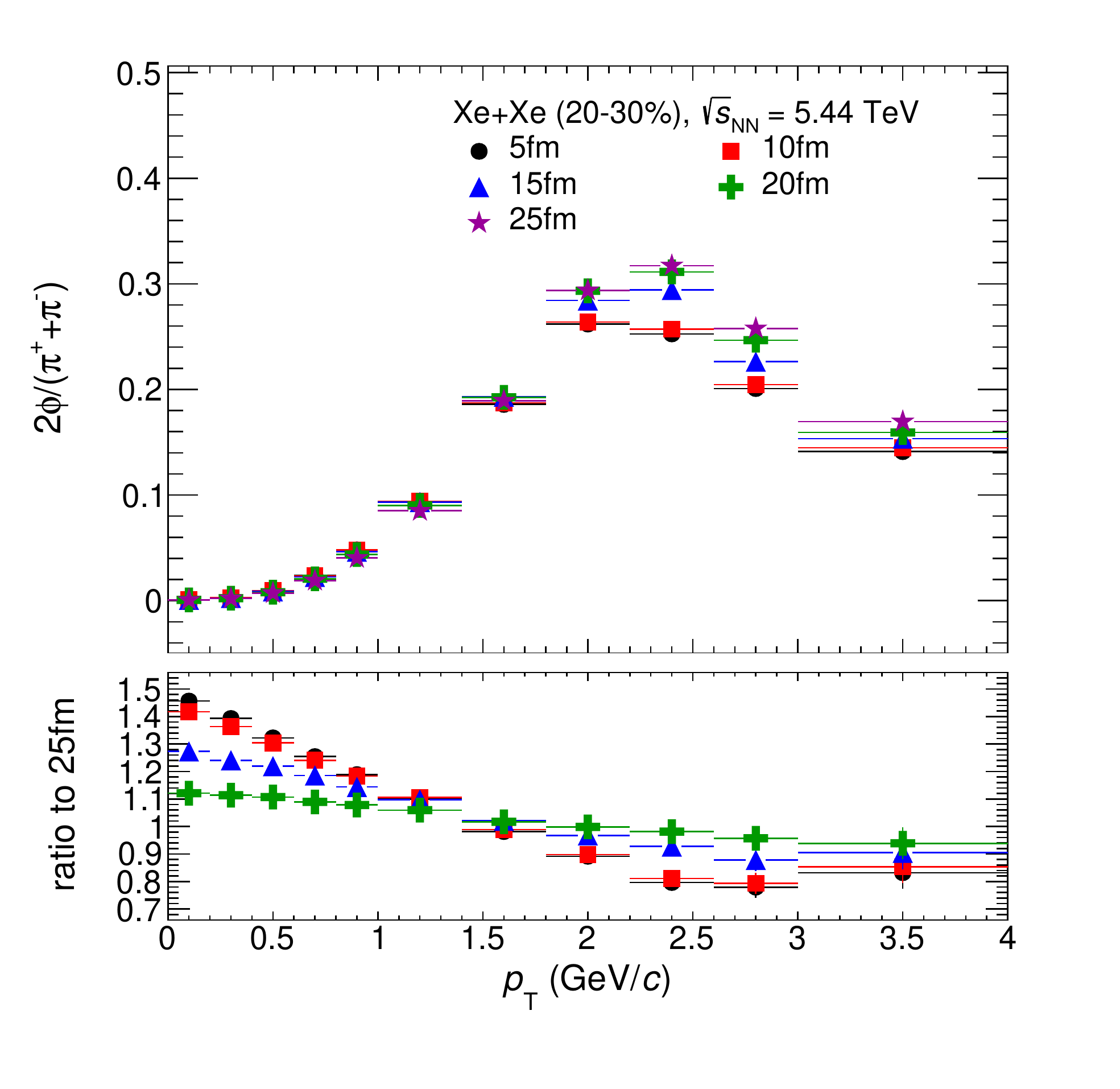}
\includegraphics[width=6cm, height=6cm]{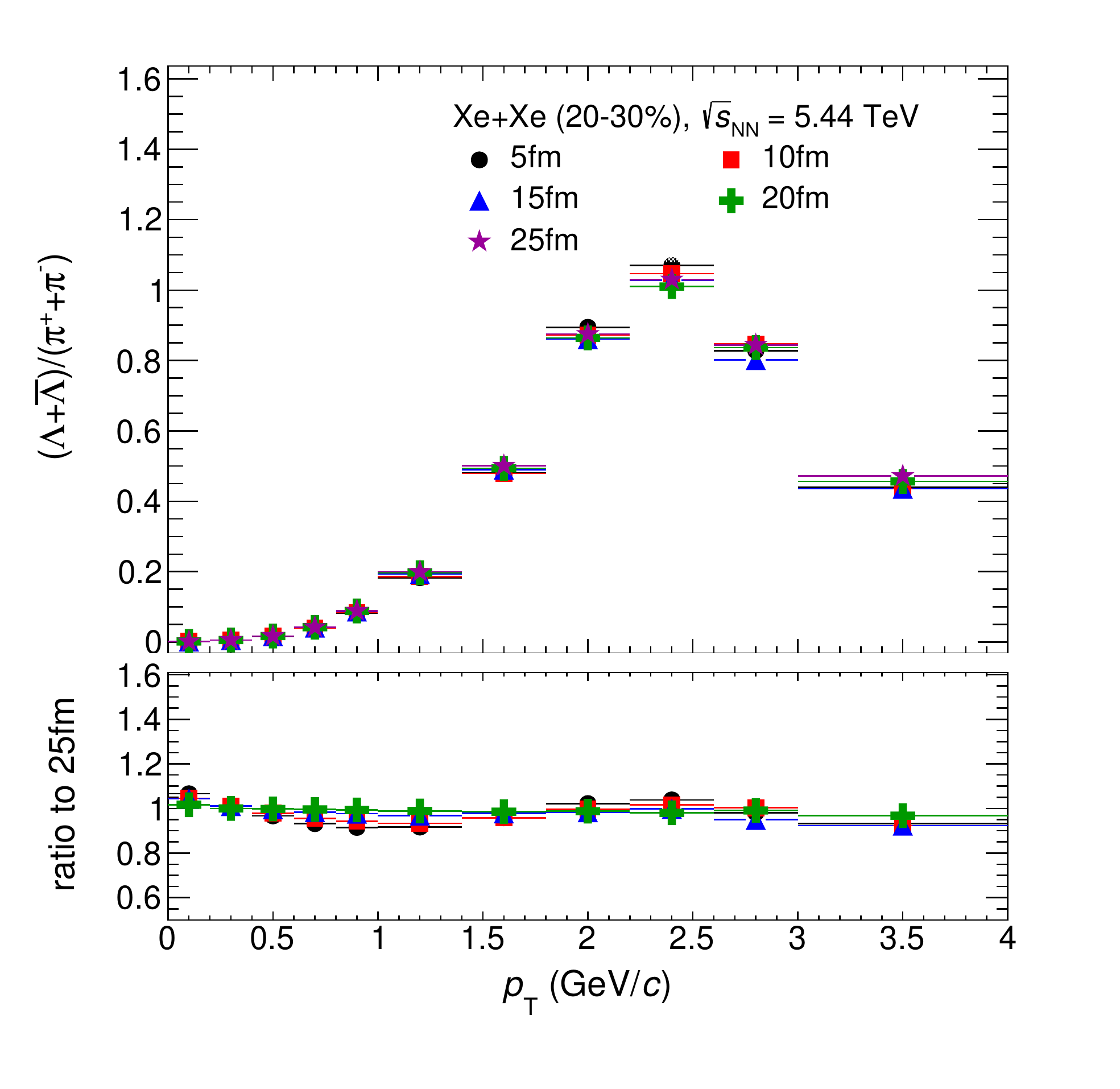}
\caption[]{(Color online) $p\rm{_T}$-differential particle ratios of $K^{\pm}$, $p(\bar{p})$, $\phi$ and $\Lambda(\bar{\Lambda})$ to $\pi^{\pm}$ in Xe+Xe collisions at $\sqrt{s_{\rm NN}}$ = 5.44 TeV for 20--30 $\%$ centrality class. Different symbols show various hadron cascade-times. The vertical lines on the data points are the statistical uncertainties.}
\label{pi_ratio1}
\end{figure*}

\section{Formalism}
\label{formalism}
\subsection*{A Mutli-Phase Transport (AMPT) model}
\label{ampt}

In this study, we have used A Multiphase Transport (AMPT) model which is a hybrid transport model \cite{AMPT1,AMPT2,AMPT3}. It has four components, namely: Heavy-Ion Jet INteraction Generator (HIJING) that is used for the initialization of collisions \cite{ampthijing}, Zhang's Parton Cascade model (ZPC) that helps in parton transport after initialization, Lund string fragmentation model or quark coalescence model to carry-out the hadronization mechanism and finally A Relativistic Transport (ART) model to complete the hadron transport. The differential cross-section of the produced minijet particles in p+p collisions is calculated in the HIJING model which is given by,

\begin{eqnarray}
\frac{d\sigma}{dp_T^2\,dy_1\,dy_2}=&K&\,\sum_{a,b}x_1f_a(x_1,p_{T1}^2)\,x_2 f_b(x_2,p_{T2}^2)\nonumber\\
&\times&\frac{d\hat{\sigma}_{ab}}{d\hat{t}}\,,
\end{eqnarray}

where $\sigma$ and $\hat{t}$ are the produced particles cross-section and the momentum transfer during partonic interactions in p+p collisions, respectively.  $x_i$'s and $f(x, p_T^2)$'s represent the momentum fraction of the mother protons which are carried by interacting partons and the parton density functions (PDF) accordingly. With the help of parametrized shadowing function and nuclear overlap function of the in-built Glauber model, the partons produced in the p+p collisions are transformed to A+A and p+A collisions within HIJING. We use Woods-Saxon (WS)~\cite{Loizides:2017ack} distribution to define the distribution of nucleons inside the spherical nuclei like Au and Pb. Whereas, for the deformed nuclei like Xenon, we have used a modified Woods-Saxon (MWS) density distribution in the HIJING that includes deformation parameter, $\beta_n$, along with spherical harmonics, $Y_{nl}(\theta)$, in the WS function~\cite{mws_1,mws_2, mws_3,mws_4,mws_5}. Nucleon density is usually written as a three parameter Fermi distribution \cite{density_fermitype}, which is defined as,

\begin{equation}
\rho (r) = \rho_0 \bigg[\frac{1+w(r/R)^{2}}{1+exp[(r-R)/a]}\bigg]\,.\\
\label{eqn1}
\end{equation}
Here $\rho_0$ is the nuclear matter density in the centre of the nucleus, $w$ is the deviation from a smooth spherical surface and $a$ is the skin depth or surface thickness. The parameter, $R$, is the radius of the nucleus from its centre and $r$, is a position variable and distance of any point from the centre of the nucleus. Spherical nuclei are considered to have a uniform distribution of nucleons within their spherical volume and a smooth surface for which the value of $w$ becomes zero. Hence, the reduced nucleonic density function \cite{wdsx} for a spherical nucleus can be written as,

\begin{equation}
\rho (r) = \frac{\rho_0}{1+exp[(r-R)/a]}\,.\\
\label{eqn2}
\end{equation}

The radius parameter, $R$, can be modified for an axially symmetric or prolate deformed nucleus like Xe. The modified Woods-Saxon nuclear radius \cite{wdsx_deform} is given by,

\begin{equation}
R_{A\Theta} = R[1+ \beta_2 Y_{20}(\theta)+\beta_4 Y_{40}(\theta) ],
\label{eqn3}
\end{equation}
where the $\beta_i$'s are deformation parameters. In case of Xenon nucleus, the deformation parameters, $\displaystyle\beta_2$= 0.162 and $\displaystyle\beta_4$= -0.003, which are taken from Ref. \cite{atomic_data_table,Giacalone:2017dud}.
The spherical harmonics, $Y_{20}$, and $Y_{40}$ are given by~\cite{speherical_harmonic}, \\
\begin{eqnarray}
Y_{20}(\theta) &=& \frac{1}{4} \sqrt{\frac{5}{\pi}}(3 \ cos^2\theta -1)\nonumber\\
Y_{40}(\theta) &=& \frac{3}{16\sqrt\pi}(35 \ cos^4\theta -30 \ cos^2\theta+3)\,.
 \end{eqnarray}
 
\begin{figure*}[hbt!]
\includegraphics[width=6cm, height=6cm]{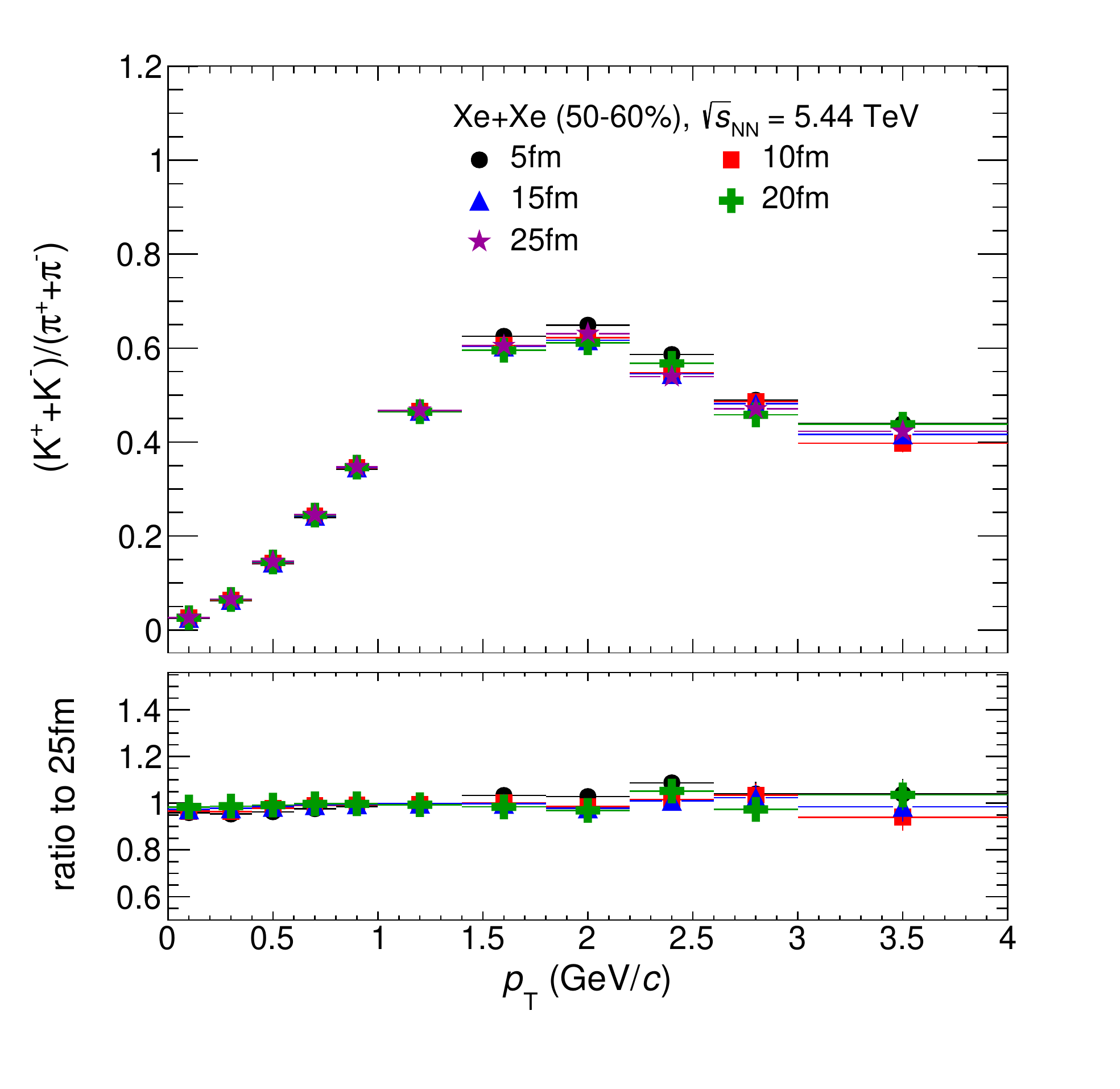}
\includegraphics[width=6cm, height=6cm]{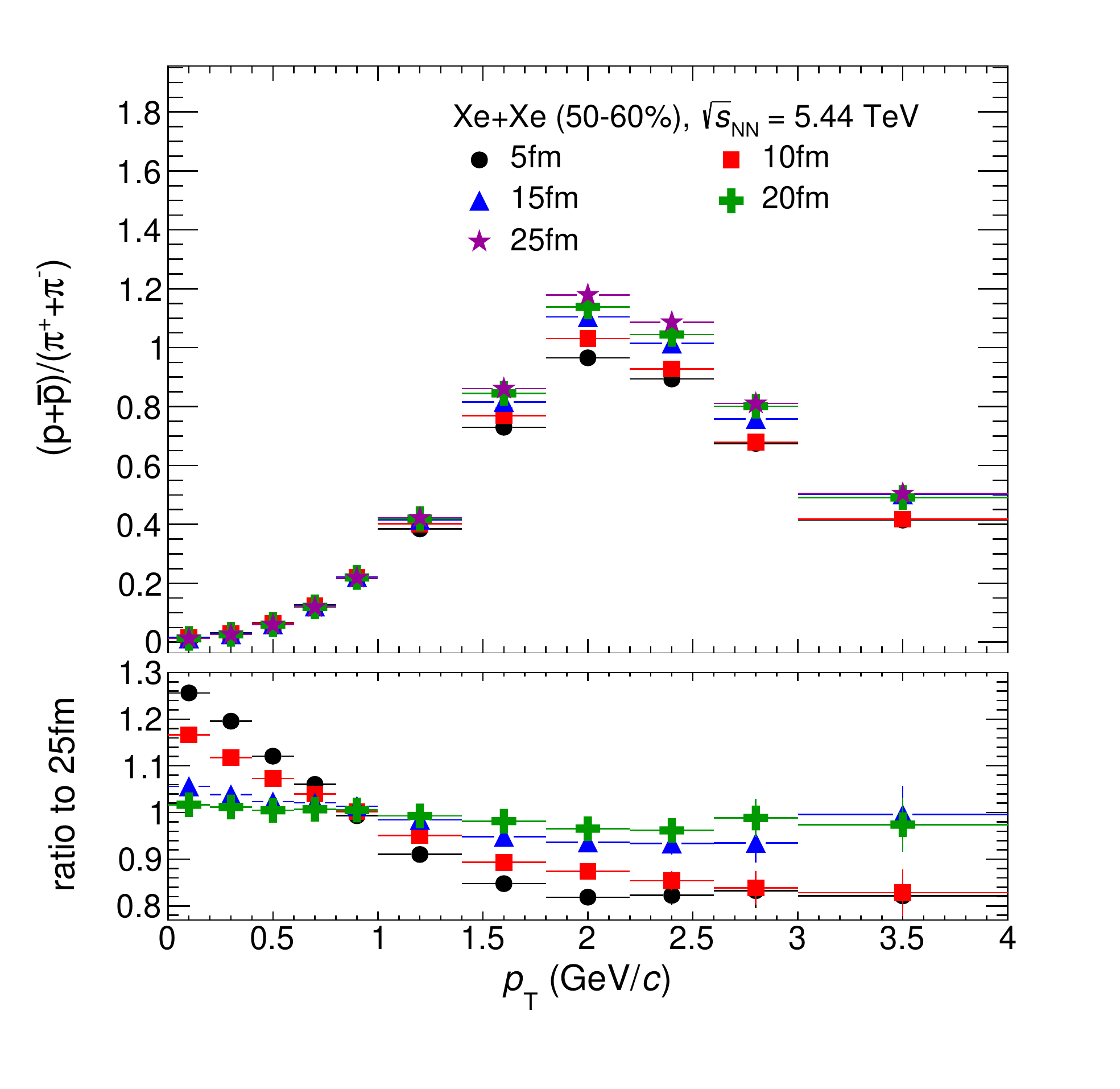}
\includegraphics[width=6cm, height=6cm]{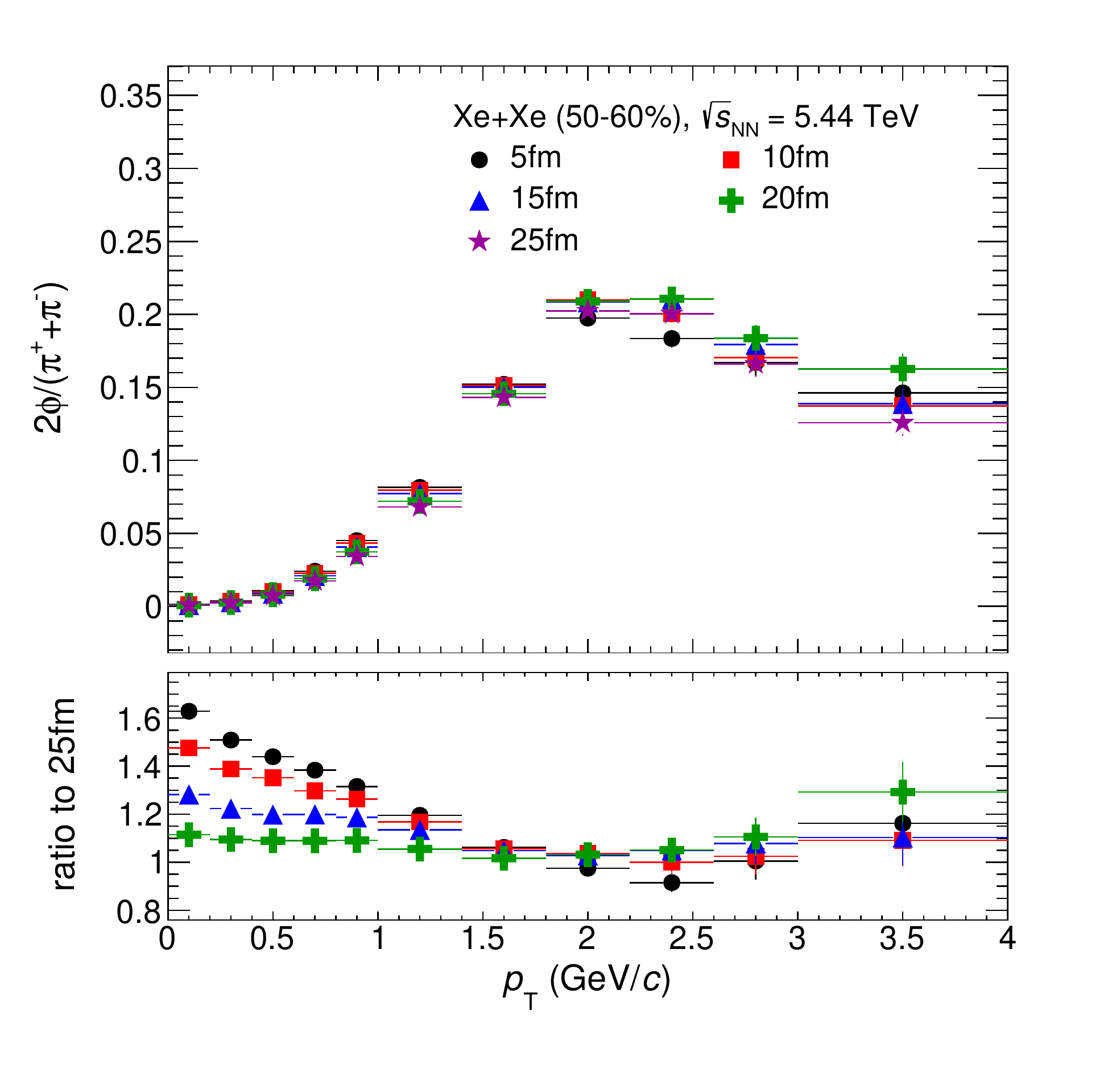}
\includegraphics[width=6cm, height=6cm]{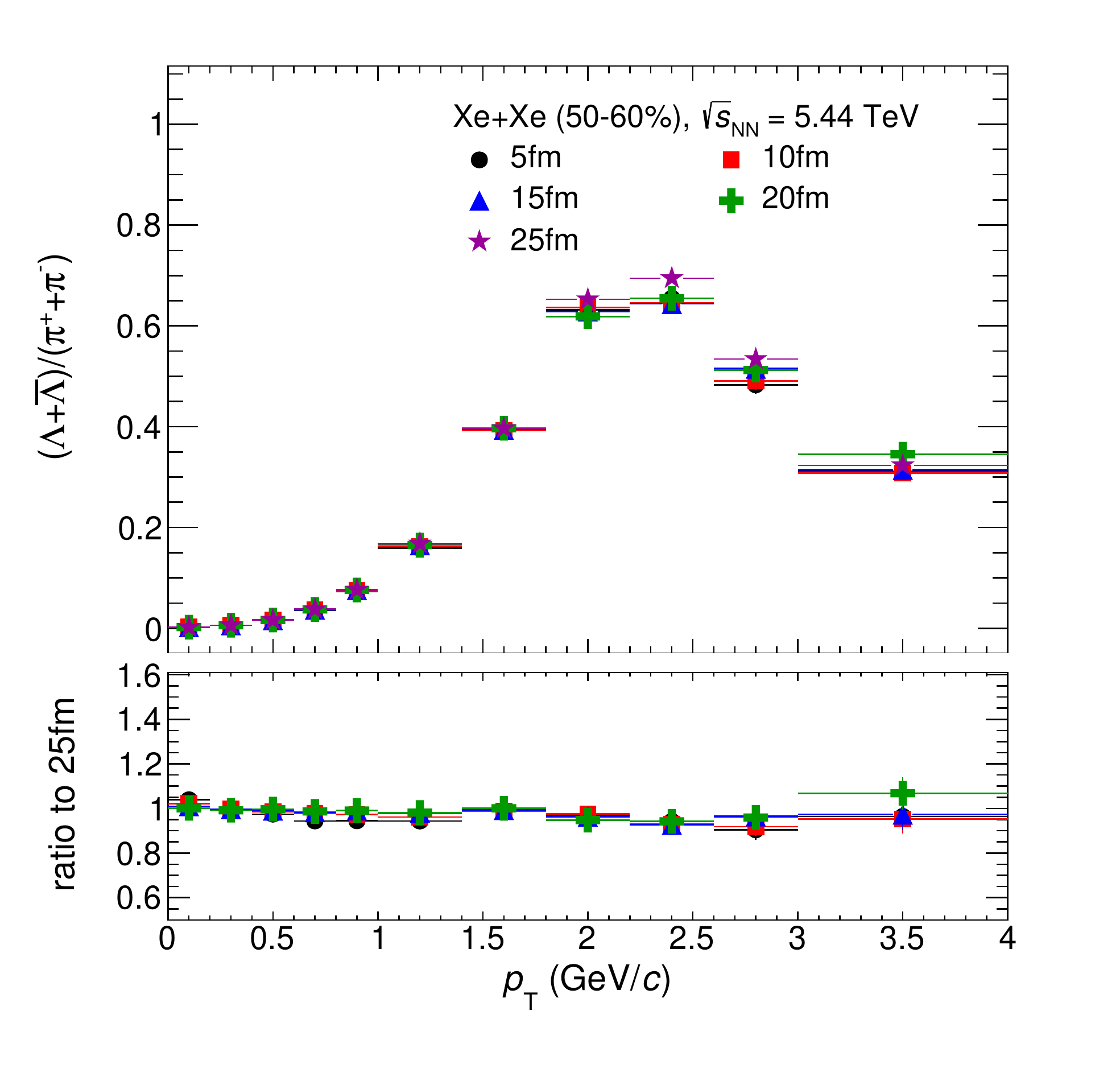}
\caption[]{(Color online) $p\rm{_T}$-differential particle ratios of $K^{\pm}$, $p(\bar{p})$, $\phi$ and $\Lambda(\bar{\Lambda})$ to $\pi^{\pm}$  in Xe+Xe collisions at $\sqrt{s_{\rm NN}}$ = 5.44 TeV for 50--60$\%$ centrality class. Different symbols show various hadron cascade-times. The vertical lines on the data points are the statistical uncertainties.}
\label{pi_ratio2}
\end{figure*}

The positions of nucleons within the distribution, $\rho(r)$, are illustrated using the volume element $r^2 sin \theta \ dr \ d\theta \ d\phi$ \cite{thesis_chris, ptribedy}. For random orientation of nuclei, position configurations are measured using both polar angle, (angle between major axis and beam axis) $\Theta$ in $[0, \pi]$ and azimuthal angle, (angle between major axis and impact parameter) $\Phi$ within limits $[0, 2\pi]$. Both target and projectile nuclei are rotated event-by-event in $\Theta$ and $\Phi$ space. 
In this paper, calculations have been done only with random orientation which means, unpolarized and averaged value over random $\Theta$ and $\Phi$~\cite{define_config}.  

The produced particles are then passed through the ZPC \cite{amptzpc}, that transports the quarks and gluons using Boltzmann transport equation which is given by,
\begin{eqnarray}
p^{\mu}\partial_{\mu}f(x,p,t)=C[f]
\end{eqnarray}
The leading order equation showing interactions among partons is approximately given by,
\begin{equation}
\frac{d\hat{\sigma}_{gg}}{d\hat{t}}\approx \frac{9\pi\alpha_s^2}{2(\hat{t}-\mu^2)^2}\,.
\end{equation}
Where $\sigma_{gg}$ is the gluon scattering cross-section and $\alpha_s$ is the strong coupling constant. Here, $\mu^2$ is the cutoff used to avoid infrared divergences that may occur if the momentum transfer, $\hat{t}$, goes to zero during scattering. 

In the AMPT-SM version, colored strings melt into low momentum partons which take place at the start of the ZPC and are calculated using the Lund FRITIOF model of HIJING. These resulting partons undergo multiple scattering when any two partons are within minimum separation which is given by $\displaystyle d\,\leq\,\sqrt{\sigma/\pi}$, where $\sigma$ is the scattering cross-section of the partons. Further, using the coalescence mechanism, the transported partons are hadronized \cite{amptreco, ampthadron2, ampthadron3}. Finally, the produced hadrons undergo evolution through the ART model, where interactions take place among meson-meson, baryon-baryon and meson-baryon before we get the final state hadrons \cite{amptart1, amptart2}. For our work, we have used the AMPT version 2.26t7 (released: 28/10/2016) where we have fixed the partonic scattering cross-section ($\sigma_{gg}$) = 10 mb and the Lund string fragmentation parameters $a$ and $b$ are set to their default values of 2.2 and 0.5/GeV$^2$, respectively. It is worthwhile to note that we have kept the hadron level decay flagged as off for $\phi$ and $\rm{K}_{s}^0$ to study these particles in the final state. In the AMPT, the hadron cascade-time is controlled by the parameter named NTMAX and DT. By varying DT, we have generated the data in the AMPT-SM for the hadron cascade-times from 5 to 25 fm/$c$ within a time interval of 5 fm/$c$. Since the hadronic phase lifetime for the default version of the AMPT is 30 fm and the lower limit obtained from the study of re-scattering effect using resonances with experimental data is $\sim$ 2--4 fm \cite{ALICE:2019xyr,XX,Sahu:2019tch}, we have taken such an interval of hadron cascade-time for our studies. 

\section{Results and discussions}
\label{ResDis}

In this work, we have investigated  $p\rm{_T}$-differential particle ratios, $p\rm{_T}$-differential and $p\rm{_T}$-integrated elliptic flow for identified and charged particles for different centralities, respectively. The results are obtained for Xe+Xe collision at $\sqrt{s_{\rm NN}}$ = 5.44 TeV by analysing the AMPT-SM generated events in which we have taken care of the deformation effect of Xe-nuclei. Events are generated for a wide range of hadronic cascade-time ranging from 5 to 25 fm/$c$ within a time interval of 5 fm/$c$. Here, we have studied the $p\rm{_T}$-differential identified particle ratios with respect to ($\pi^{+}+\pi^{-}$) and (p+$\bar {\rm{p}}$). Furthermore, we have tried to see the quark participant scaled elliptic flow for identified particles such as ($\pi^{+}+\pi^{-}$), ($K^{+}+K^{-}$), (p+$\bar {\rm{p}}$), $\phi$ and ($\Lambda+\bar {\rm{\Lambda}}$). Hadronic cascade-time can be sensitive to the $p\rm{_T}$-differential identified particle ratios depending upon their scattering cross-section in the hadronic phase and the lifetime of the hadronic phase. Further, the multiple scattering of the final state particles may modify their azimuthal direction and transverse momentum which will be reflected on the $p\rm{_T}$-differential and $p\rm{_T}$-integrated elliptic flow.  Hence, our primary goal is to quantify the effect of hadronic cascade-time ($\tau_{HC}$) on the observables like $p\rm{_T}$-differential particle ratios and elliptic flow in the Xe+Xe collision system. This study will enrich our understanding of particle production dynamics and the effect of final state hadronic cascade-time in view of the finite hadronic phase lifetime at the LHC energies, even in small collision systems.

\subsection{Identified $p\rm{_T}$-differential particle ratios}
\begin{figure}[hbt!]
\includegraphics[scale=0.42]{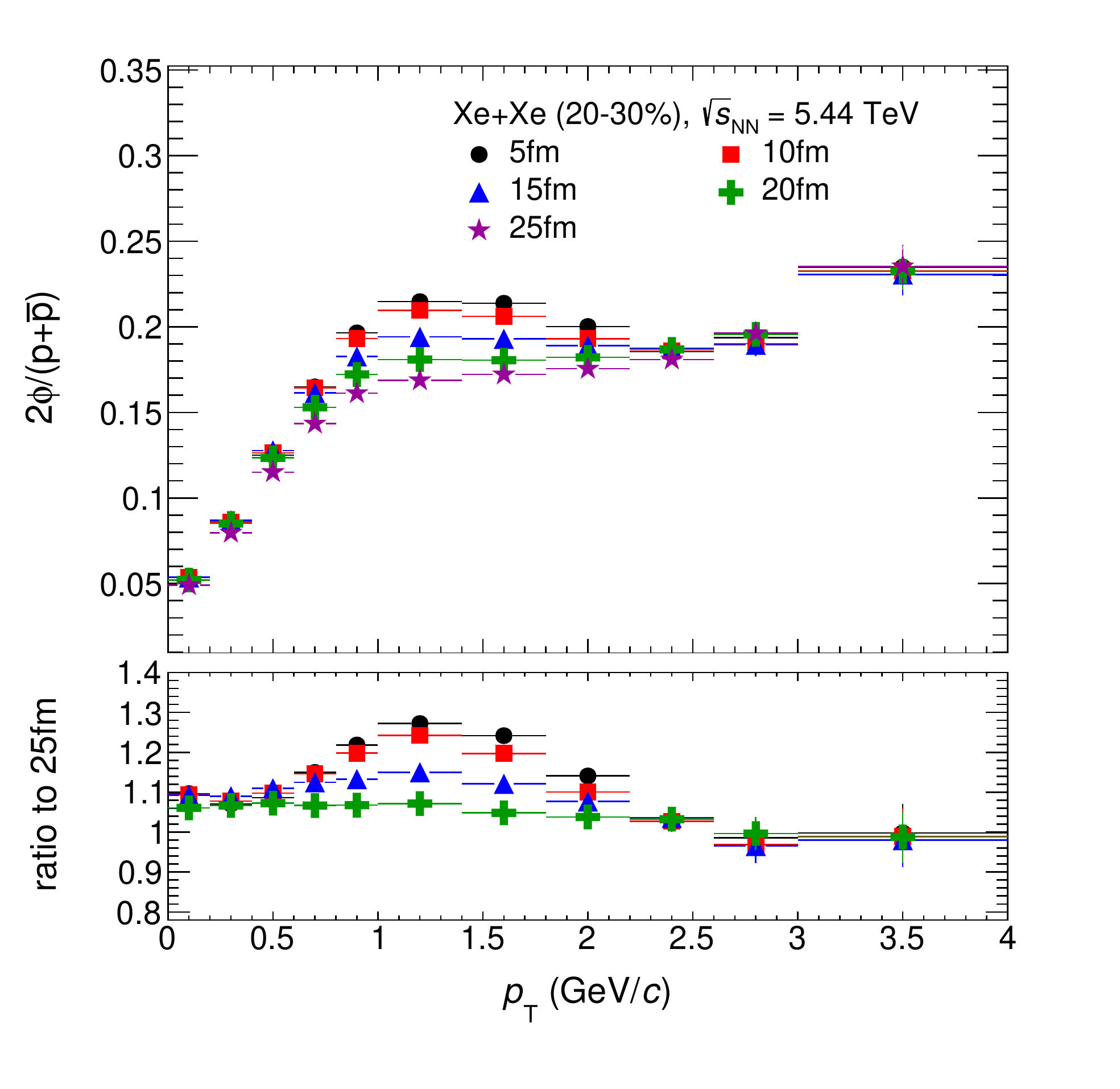}
\includegraphics[scale=0.42]{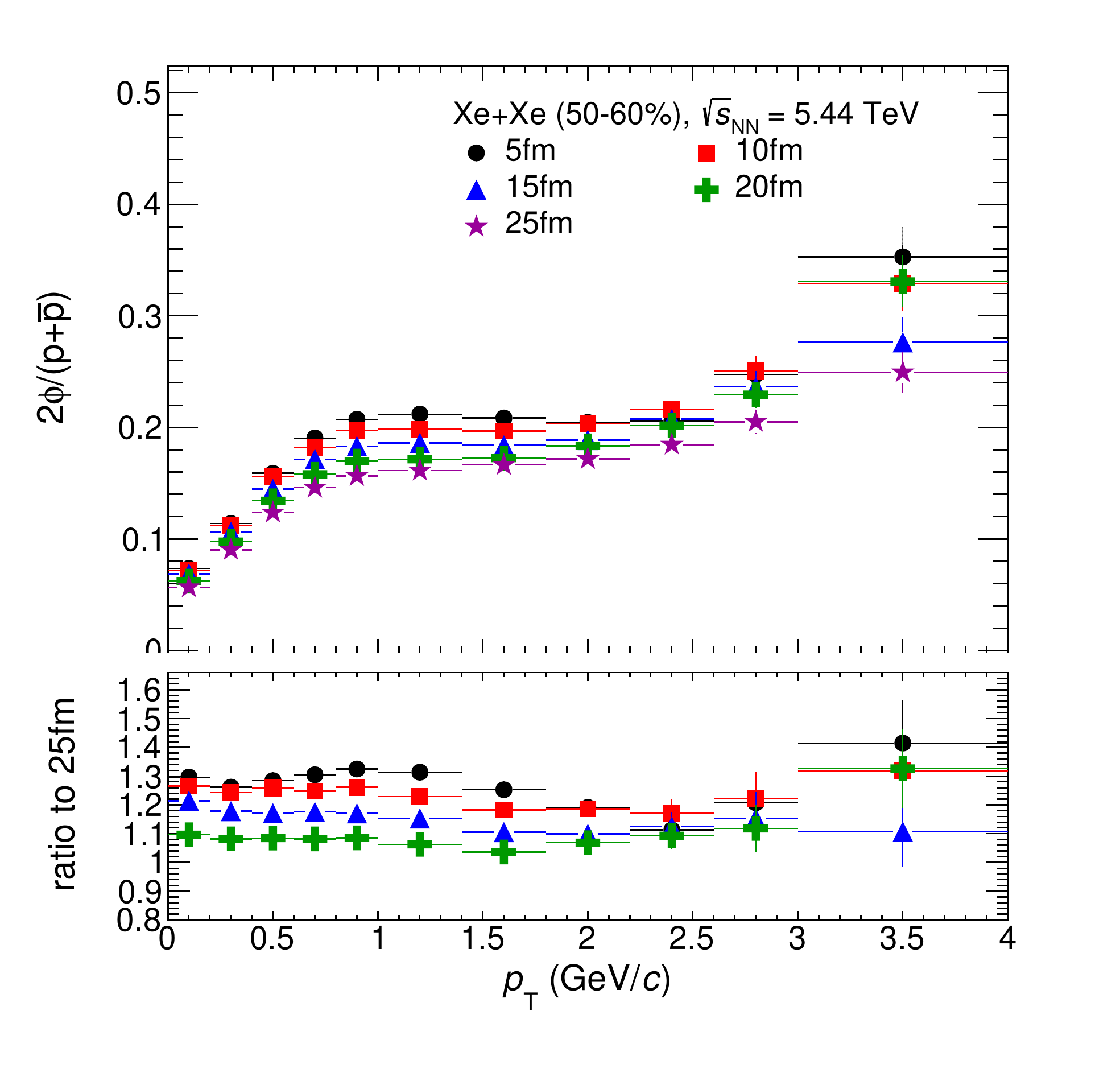}
\caption[]{(Color online) $p\rm{_T}$-differential particle ratios of $\phi$ to p in Xe+Xe collisions at $\sqrt{s_{\rm NN}}$ = 5.44 TeV for 20--30 $\%$ (upper) and 50--60 $\%$ (lower) centrality classes. Different symbols show various hadron cascade-time. The vertical lines in the data points are the statistical uncertainties.}
\label{P_ratio1}
\end{figure}

Figures \ref{pi_ratio1} and \ref{pi_ratio2} show the ratio to pion in 20--30$\%$ and 50--60$\%$ centrality classes for the particles like $K^{+}+K^{-}$, p+$\bar {\rm{p}}$, $\phi$ and $\Lambda+\bar {\rm{\Lambda}}$ as a function of transverse momentum in different hadronic cascade-time. The production rate of these identified particles as compared to pion increases with $p\rm{_T}$ which attain a maximum value in the intermediate $p\rm{_T}$ (2--3 GeV/$c$). This similar behaviour is also observed for all the hadronic cascade-time. At lower transverse momentum region ($p\rm{_T}~< 1~{\rm GeV}/c$), we see a significant effect of the hadronic cascade-time on $\phi$/$\pi$ and p/$\pi$ which have a maximum deviation of $\sim$ 40$\%$ for $\tau_{HC}$ = 5 fm/$c$ with respect to 25 fm/$c$. However, this trend reverses for $p\rm{_T}~> 1~{\rm GeV}/c$. Moreover, the hadronic cascade-time has less effect on the strange particle ratios like K/$\pi$ and $\Lambda$/$\pi$. For 50--60$\%$ centrality class, the deviation for $\phi$/$\pi$ ratio is higher at lower $p\rm{_T}$ region as compared to 20--30$\%$ centrality class.

Furthermore, we have investigated the effect of hadronic cascade-time on the $\phi/p$ $p\rm{_T}$-differential ratio, which has a nearly similar mass. Figure \ref{P_ratio1} represents $\phi$/p ratio as a function of $p\rm{_T}$  in 20--30$\%$ and 50--60$\%$ centrality classes for different hadron cascade-times. The ratio increases with $p\rm{_T}$ and starts saturating after $p\rm{_T}$ = 1 GeV/$c$ for both the centrality classes that further enhances at high-$p_{\rm T}$. A similar trend is followed by all the hadronic cascade-times. From the lower panel of the figure, we observe a clear dependence on the $\tau_{HC}$ especially for 50--60$\%$ centrality class which gradually increases as we decrease the value of $\tau_{HC}$ and shows a maximum deviation $\sim$ 30$\%$ for $\tau_{HC}$ = 5 fm/$c$ as compared to 25 fm/$c$.

This study signifies the importance of different processes in the hadronic phase and their impact on the $p\rm{_T}$-differential identified particle ratios. This motivates us to further investigate the effect of hadronic cascade-time on one of the important observables known as elliptic flow and the results will be discussed in the next section, \ref{elliptic}.

\subsection{Elliptic flow ($v_2$)}
\label{elliptic} 
\begin{figure}[ht!]
\includegraphics[scale=0.4]{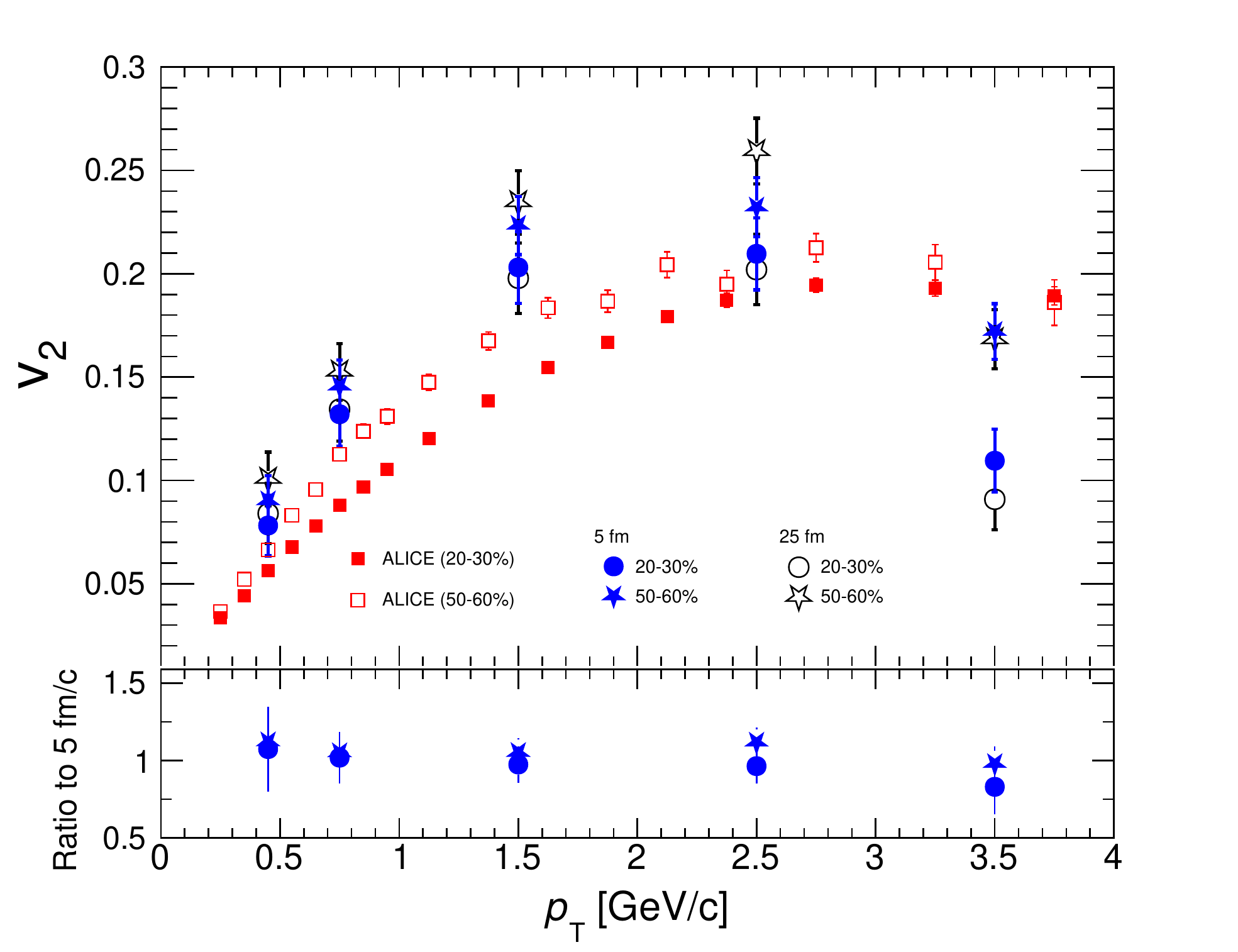}
\caption[]{(Color online) $p\rm{_T}$-differential elliptic flow of charged particles in Xe+Xe collisions at $\sqrt{s_{\rm NN}}$ = 5.44 TeV for 20-30$\%$ and 50-60$\%$ centrality classes. Different symbols show various hadron cascade-times. The results are compared with the ALICE data \cite{Acharya:2018ihu}.}
\label{v2_comp}
\end{figure}

To be inline with the experimental procedure of estimating elliptic flow, in this study, we have used the two-particle correlation method for a direct comparison of AMPT data. The details of the two-particle correlation method are described elsewhere \cite{Mallick:2020ium}. For Pb+Pb collisions at $\sqrt{s_{\rm NN}}$ = 2.76 TeV, a two-particle
correlation method was also used to predict $v_{\rm 2}$ for identified hadrons using AMPT \cite{Xu:2011fi}. In addition, the same method was also successfully used for estimating higher-order anisotropic flows, as well as dihadron correlations in Pb+Pb collisions at $\sqrt{s_{\rm NN}}$ = 2.76 TeV using the AMPT model \cite{Xu:2011jm}. The two-particle correlation method has the added edge of construction with a proper pseudo-rapidity cut. This removes substantial residual non-flow effects in the elliptic flow. Non-flow effects are azimuthal correlations, which usually occur from jets and resonance decays, and are not associated with the symmetry planes. Figure \ref{v2_comp} shows the charged particle elliptic flow as a function of $p\rm{_T}$ for both 20--30$\%$ and 50--60$\%$ centrality classes. Here, we have compared the $p\rm{_T}$-differential elliptic flow generated for the two extreme cases, $\tau_{HC}$ = 5 fm/$c$ and 25 fm/$c$ for both the centrality classes. With different centrality classes and cascade-times, the elliptic flow increase with $p\rm{_T}$ and obtain a maximum value for $p\rm{_T}$ around 2--2.5 GeV/$c$. We observe higher $v_{\rm 2}$ for the peripheral collisions (50--60$\%$) as compared the semi-central collisions (20--30$\%$). The results are compared with the $p\rm{_T}$-differential elliptic flow of the charged particles obtained from ALICE for both  20--30$\%$ and 50--60$\%$ centrality classes. In both the cases, the trend is qualitatively explaining the ALICE data and a higher elliptic flow is observed for 50--60 $\%$ centrality class, which is in line with the experimental observations. 

 This can be understood from the effect of initial momentum anisotropic due to the geometry of the overlapping region generated after the collisions which are higher for the peripheral collisions. While considering different hadronic cascade-times, we observe a similar behavior of the $p\rm{_T}$-differential elliptic flow in both 20--30$\%$ and 50--60$\%$.  For both the centralities, with higher cascade-time, the azimuthal distribution of the final state particles is more anisotropic for high and very low-$p_{\rm T}$ region. However, in the intermediate-$p_{\rm T}$ region seems to be not affected by the hadronic cascade-time significantly.
 
\begin{figure}[ht!]
\includegraphics[scale=0.42]{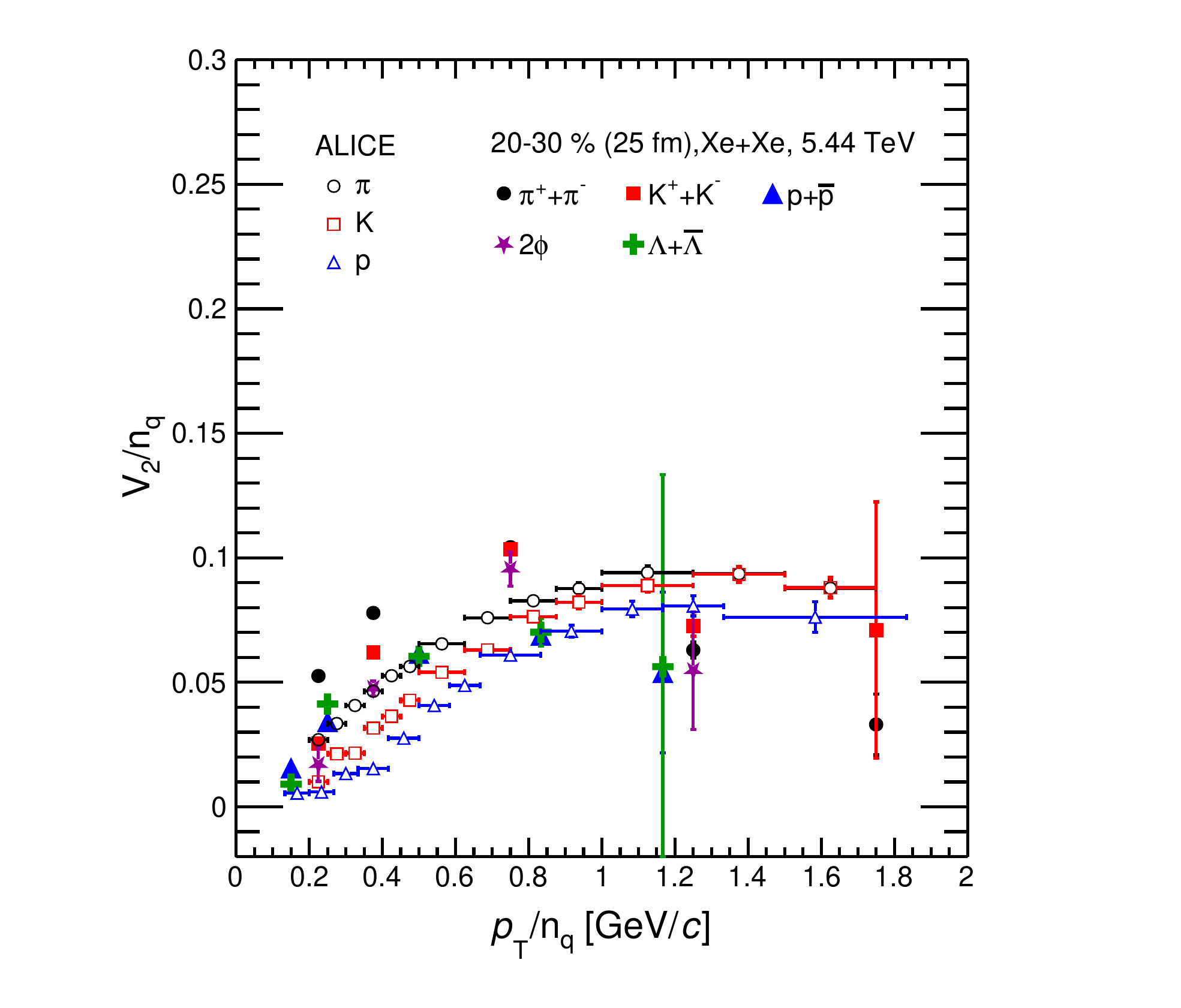}
\includegraphics[scale=0.42]{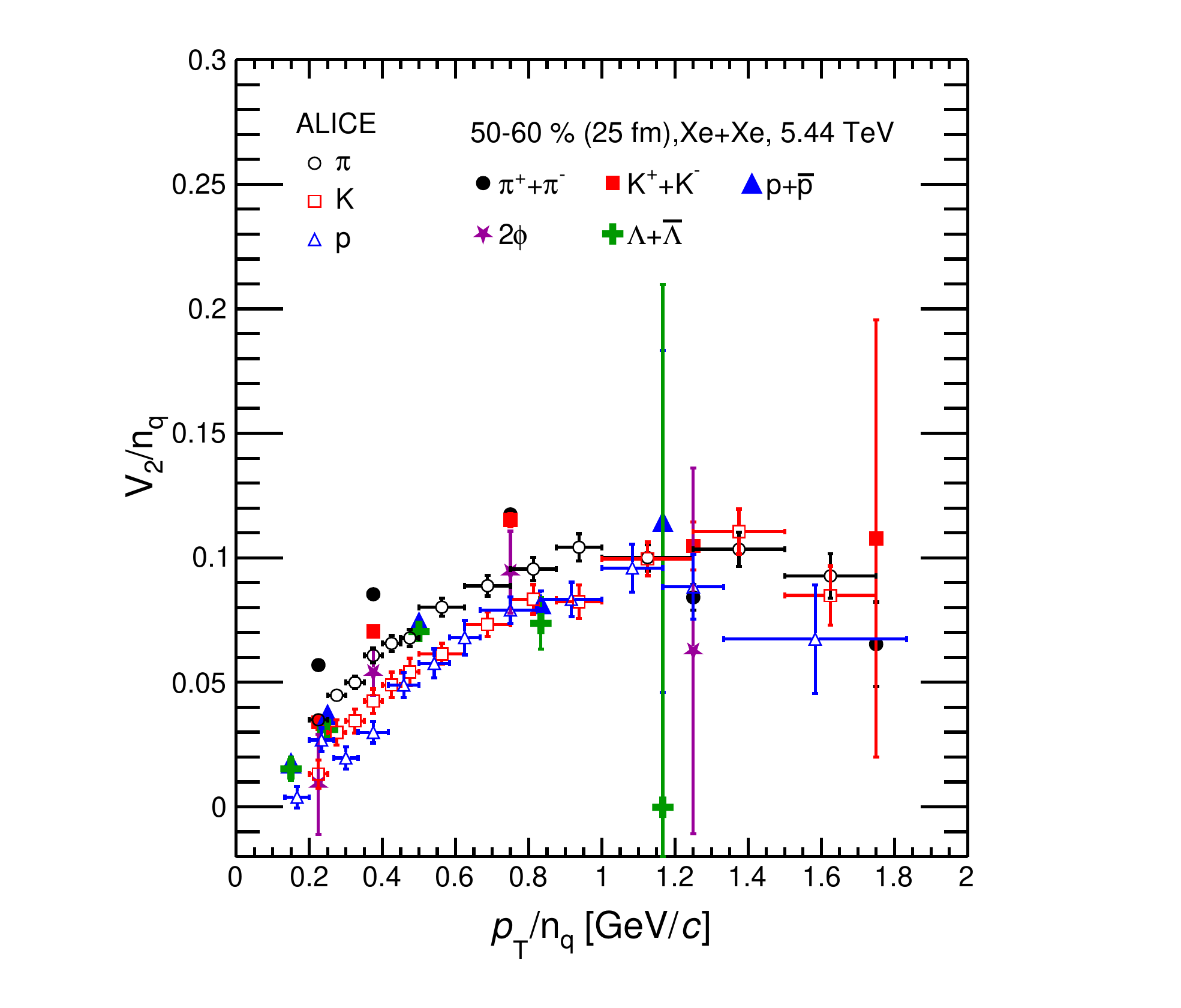}
\caption[]{(Color online) $n_{\rm q}$-scaled $p\rm{_T}$-differential elliptic flow of the identified charged particles in Xe+Xe collisions at $\sqrt{s_{\rm NN}}$ = 5.44 TeV for 20--30$\%$ (top) and 50--60$\%$ (bottom) centrality classes for hadron cascade-time of 25 fm/c. Different symbols show various identified hadrons. The results are compared with the ALICE data \cite{ALICE:2021ibz}.}
\label{v2_ptdiff}
\end{figure}

The particle production via quark coalescence can be verified by studying the elliptic flow of baryons and mesons at intermediate $p\rm{_T}$ after scaling both $p\rm{_T}$ and $v_{\rm 2}$ by the number of constituent quarks ($n_{\rm q}$). In AMPT with the string melting mode, mesons and baryons are formed when a quark and antiquark pair and three quarks come close in phase space, respectively. In such conditions, the elliptic flow for baryons is higher at the intermediate $p\rm{_T}$ range compared to mesons due to the recombination effect. Figure \ref{v2_ptdiff} represents the constituent quark scaled $v_2$ as a function of $p\rm{_T}$/$n_{\rm q}$. For this study, the elliptic flow and $p\rm{_T}$ are scaled by 2 for mesons like pions, kaons and $\phi$, while the same are scaled by 3 for baryons such as protons and $\Lambda$. We can clearly see the violation of the scaling behavior in both 20--30$\%$ and 50--60$\%$ centrality classes at $\tau_{HC}$ = 25 fm/$c$, which are in line with the deviations that are already reported in ALICE \cite{ALICE:2021ibz} and  several experimental studies performed at LHC energies \cite{Acharya:2018zuq}. We have explicitly checked that this violation is also observed for $\tau_{HC}$ = 5 fm/$c$, indicating the fact that
hadron cascade-time has no role on the quark-participant scaling violation in the elliptic flow.

\begin{figure}[ht!]
\includegraphics[scale=0.42]{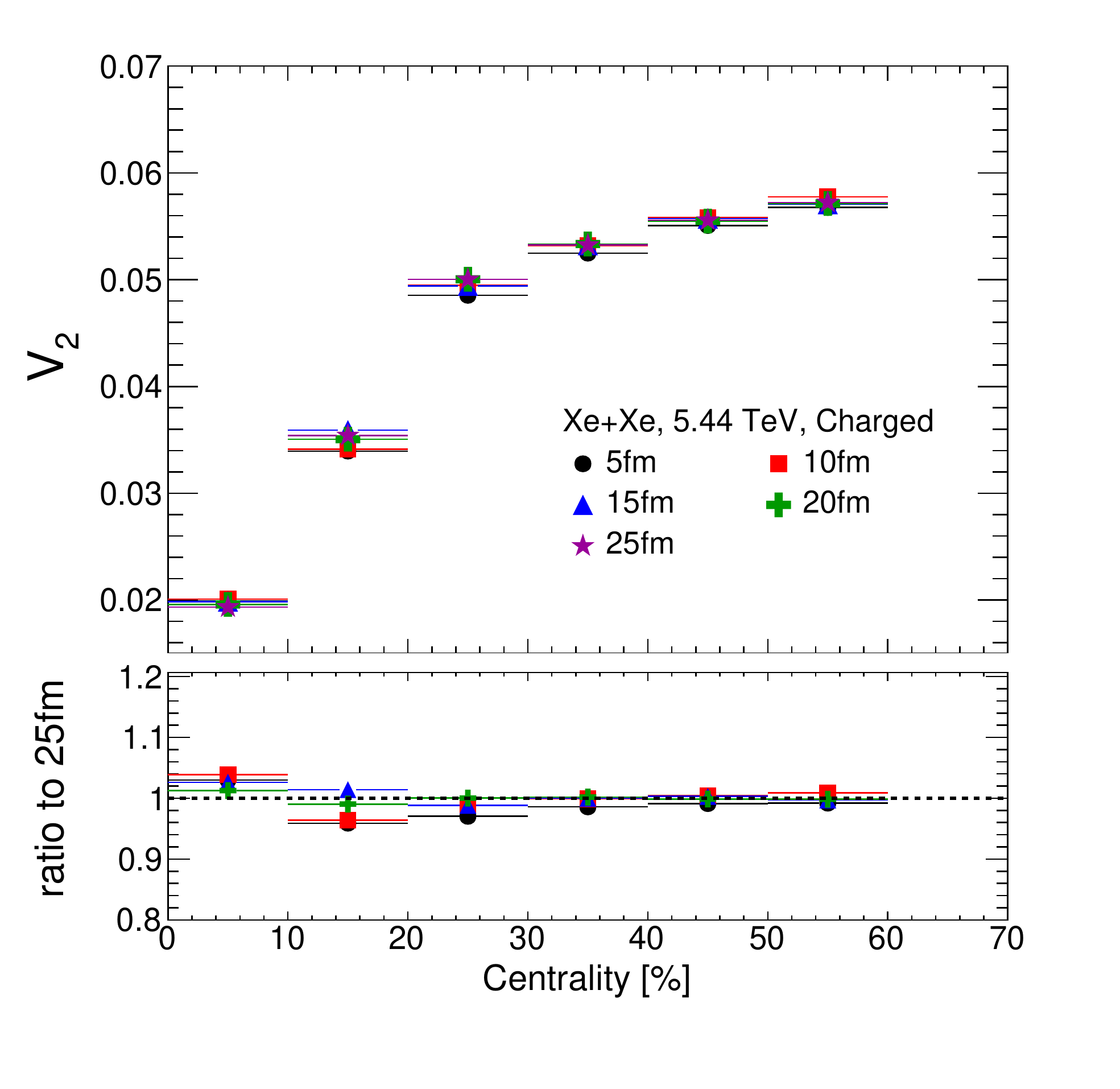}
\caption[]{(Color online) $p\rm{_T}$-integrated elliptic flow of the charged particles in Xe+Xe collisions at $\sqrt{s_{\rm NN}}$ = 5.44 TeV for 0-10$\%$, 10-20$\%$, 20-30$\%$, 30-40$\%$, 40-50$\%$ and 50-60$\%$ centrality classes. Different symbols show various hadron cascade-times.}
\label{v2_ptInt}
\end{figure}

Furthermore, to investigate the effect of hadronic cascade-time on the bulk part of the system with centrality, we have studied the $p\rm{_T}$-integrated elliptic flow of the charged particles. Inspired by the ALICE kinematic acceptance, we have considered the charged tracks with $p\rm{_T}$ lying between 0.2 to 3 GeV/$c$. Figure \ref{v2_ptInt} shows the $p\rm{_T}$-integrated elliptic flow as a function of centrality for different values of $\tau_{\rm HC}$. While going from central to peripheral collisions, the $p\rm{_T}$-integrated $v_2$ increases for all the hadronic cascade-times. To quantify the dependence on hadronic cascade-time, we have shown the ratio with respect to 25 fm/$c$ in the lower panel of Figure \ref{v2_ptInt}. We do not observe significant dependence of the $p\rm{_T}$-integrated $v_2$ on the hadronic cascade-time, though we see a clear dependence on hadronic cascade-time in case of the $p\rm{_T}$-differential $v_2$.

\section{Summary}
\label{summary}

In this study, we have explored the effect of hadronic cascade-time ($\tau_{\rm HC}$) on the $p\rm{_T}$-differential identified particle ratios, $p\rm{_T}$-differential and $p\rm{_T}$-integrated elliptic flow in Xe+Xe collision at $\sqrt{s_{\rm {NN}}}$ = 5.44 TeV using the data generated from the AMPT-SM version. In the AMPT-SM, the participation of soft partons originating from the string melting process in partonic scatterings results in an increased elliptic flow as compared to the default AMPT.  In addition, it allows modifying the final state hadronic phase lifetime by varying the number of time steps in a hadron cascade and/or the length of the time step. The significance of the hadronic cascade-time on particle production depends on the scattering cross-section among the final state particles and the duration of the hadronic phase. This study has more significance in looking into the larger hadronic phase lifetime at the LHC energies. We see a significant dependence of identified particle ratios and elliptic flow on the $\tau_{\rm HC}$ when studied as a function of $p_{\rm T}$ and centrality. The important findings of this study are summarised below:

\begin{itemize}

\item Significant dependence of $p\rm{_T}$-differential particle ratios for $\phi/\pi$ and $p/\pi$ on $\tau_{\rm HC}$ is observed at low $p_{\rm T}$, which is higher for $\phi/\pi$ ratio. With higher $\tau_{\rm {HC}}$, the low $p\rm{_T}$ particles shift towards intermediate and higher $p\rm{_T}$ region due to more interactions which is reflected in the particle ratios. 

\item To cancel out the mass dependence on $\tau_{\rm HC}$, we have looked into the $\phi/p$ $p\rm{_T}$-differential particle ratio. We see a scaling of this ratio for 50-60$\%$ centrality class with $\tau_{\rm HC}$. However, for mid-central collisions, we see a significant dependence of $\tau_{\rm HC}$ on $\phi/p$ ratio at intermediate $p\rm{_T}$ region.

\item The $p\rm{_T}$-differential charged particle elliptic flow is higher for  $\tau_{\rm HC}$ = 25 fm/$c$ compared to 5 fm/$c$ at very low and high-$p_{\rm T}$ region. This indicates added anisotropy in the azimuthal distribution of the charged particles might be originating from multiple scattering in the hadronic phase with higher $\tau_{\rm HC}$.

\item In line with the experimental results obtained at the LHC energies, we do not observe any scaling behavior with the number of constituent quarks ($n_{\rm q}$) on elliptic flow. The hadron cascade-time has no effect on the quark-participant scaling violation in the
elliptic flow, which is supposed to be an initial state effect in contrast to the hadronic rescattering, which is a final state effect.

\item To see the effect of hadronic cascade-time over the bulk of the medium, we have estimated the $p\rm{_T}$-integrated charged particle elliptic flow in different centrality classes. We found the $p\rm{_T}$-integrated charged particle elliptic flow is almost independent of the hadronic cascade-time. This might be due to the compensation of anisotropy over different $p\rm{_T}$ regions.

\end{itemize}

More precisely, we observe the effect of hadron cascade-time on $p\rm{_T}$-differential identified particle ratios, $p\rm{_T}$-differential and integrated charged particle elliptic flow. We see an interplay of different hadronic phase effects such as scattering cross-sections, hadronic phase lifetime and momentum anisotropy inherited from initial collision geometry on these observables.

\section*{Acknowledgements}
RR and GSP acknowledge the financial supports by DST-INSPIRE program of Government of India. Ronald Scaria acknowledges CSIR,
Government of India for financial supports.
Raghunath Sahoo acknowledges the financial supports under the
CERN Scientific Associateship and the financial grants
under DAE-BRNS Project No. 58/14/29/2019-BRNS of
Government of India. The authors would like to acknowledge the usage of resources  of the LHC grid computing facility at VECC, Kolkata
and usage of resources of the LHC grid Tier-3 computing facility at IIT Indore. We would like to thank N. Mallick and S. Prasad for helping in the estimation of elliptic flow in the two-particle correlation method. Mr. Vikash Yadav is acknowledged for making an initial exploration of the idea used in this paper as a part of his master's thesis.

\end{document}